\documentclass{article}
\usepackage{spconf,amsmath,amssymb,graphicx}
\usepackage{color,cite,balance,url,booktabs,siunitx}

\usepackage{hyperref}
\usepackage{pifont}

\title{Hypothesis stitcher for end-to-end speaker-attributed ASR on long-form multi-talker recordings}
\name{Xuankai Chang$^{1\dagger*}$, Naoyuki Kanda$^{2*}$, Yashesh Gaur$^2$, Xiaofei Wang$^2$, Zhong Meng$^2$, Takuya Yoshioka$^2$ \thanks{$^\dagger$Work performed during internship at Microsoft.}\thanks{$^*$Equal contribution.}}
\address{
    $^1$Center for Language and Speech Processing, Johns Hopkins University, USA\\
    $^2$Microsoft Corp., USA\\
}
\begin{document}
\ninept
\maketitle
\begin{abstract}

An end-to-end (E2E) speaker-attributed automatic speech recognition (SA-ASR) model
was proposed recently to jointly perform speaker counting, speech recognition and speaker identification.
The model achieved a low speaker-attributed word error rate (SA-WER)
for monaural overlapped speech comprising an unknown number of speakers. 
However, the E2E modeling approach is  susceptible to the mismatch between the training and testing conditions. 
It has yet to be investigated 
whether the E2E SA-ASR model works well 
for recordings that are much 
longer than samples seen during training. 
In this work,
we first apply a known decoding technique that was developed to perform single-speaker ASR for long-form audio to our E2E SA-ASR task. 
Then, we propose a novel method using a sequence-to-sequence model, called hypothesis stitcher. 
The model takes multiple hypotheses obtained from short audio segments that are extracted from the original long-form input, and it then outputs a fused single hypothesis.
We propose several architectural variations of the hypothesis stitcher model
and compare them  with the conventional decoding methods.
Experiments using LibriSpeech and LibriCSS corpora
show that the proposed method significantly improves SA-WER
especially for long-form multi-talker recordings.
\end{abstract}
\section{Introduction}
\label{sec:intro}

Speaker-attributed automatic speech recognition (SA-ASR)
for overlapped speech
has long been studied 
to realize meeting transcription
\cite{fiscus2007rich,janin2003icsi,carletta2005ami}.
It requires counting the number of speakers,
transcribing utterances, 
and diarizing or identifying the speaker of 
each utterance from multi-talker recordings that may contain overlapping utterances.
 Despite the
  significant progress
  that has been made especially for multi-microphone settings 
  (e.g., \cite{kanda2018hitachi,Advances-Yoshioka2019,kanda2019guided,watanabe2020chime,medennikov2020stc}), 
 SA-ASR remains very challenging when we can only
 access monaural audio. 
One typical approach 
is a modular approach combining
individual modules, such as 
speech separation, speaker counting,
 ASR and speaker diarization/identification.
However, since different modules are designed based on
different criteria, 
the simple combination does not  necessarily result in an optimal solution for the SA-ASR task.

An end-to-end (E2E) SA-ASR model was recently proposed in 
\cite{Joint-Kanda2020}
for monaural multi-talker speech 
as a joint model of speaker counting, speech recognition, 
and speaker identification.
Unlike prior studies in joint SA-ASR modeling  
\cite{el2019joint,mao2020speech,kanda2019simultaneous}, 
it unifies the key components of SA-ASR, i.e., speaker counting, speech recognition, and speaker identification, and thus can handle speech consisting of any number of speakers. 
 The model 
greatly improved the 
speaker-attributed word error rate (SA-WER) over the modular system. 
\cite{Investigation-Kanda2020} also showed that the E2E SA-ASR model could be used 
even without prior speaker knowledge
by applying speaker counting and speaker clustering to
model's internal embeddings.

While promising results have been reported, 
it is still unclear
 whether the E2E SA-ASR model works well for audio recordings 
 that are much longer than those seen during training.
Our preliminary analysis using in-house real conversation data revealed
that a considerable number of
long segments existed even after 
applying
voice activity detection to the original long-form recordings. 
Note that,
for a conventional single-speaker  E2E ASR,
it is known that 
the long-form audio input causes  significant accuracy degradation  
due to the mismatch of the training and testing conditions 
\cite{chorowski2015attention,Comparison-Chiu2019,narayanan2019recognizing}.
To mitigate this, 
\cite{Comparison-Chiu2019} proposed an ``overlapping inference'' algorithm
in which the long-form audio was segmented 
by sliding windows with a 50\% shift
and the hypotheses from each window were fused
to generate one long hypothesis with an edit-distance-based heuristic.
Although the overlapping inference 
could be applied to multi-talker recordings, 
the effectiveness of the method could be negatively impacted by the speaker recognition errors.
In addition, the decoding cost in the overlapping inference is two times higher due to
the overlap of the adjacent window positions. 

With this as a background, 
in this paper, 
we propose a novel method using a sequence-to-sequence model, 
 called hypothesis stitcher, 
which 
 takes multiple hypotheses
 obtained from a sliding-window and outputs a 
 fused single hypothesis.
 We propose several architectural variants of the hypothesis stitcher model 
 and compare them with the conventional decoding methods. 
 Note that, while we propose and evaluate
 the hypothesis stitcher in the context of
 SA-ASR, the idea is directly applicable to
  a standard ASR task.
 In our evaluation using the LibriSpeech \cite{Librispeech-Panayotov2015} and LibriCSS \cite{Continuous-Chen2020} corpora, 
 we show 
 that the proposed method significantly improves SA-WER 
over the previous methods.
We also show that the hypothesis stitcher
can work with a sliding window with less than 50\% overlap and thereby reduce 
the decoding cost.

\section{Review of relevant techniques} 

In this section, we review relevant techniques 
to the proposed method.
Due to the page limitation, we only describe the overview of each technique.
Refer to the original papers for further details.
\subsection{E2E SA-ASR}
\label{sec:sa-asr}

The E2E SA-ASR was proposed as a joint model of speaker counting, speech recognition and speaker identification for monaural overlapped audio \cite{Joint-Kanda2020}.
The inputs to the model are acoustic features $\mathbf{X}=\{ \mathbf{x}_1, \dots, \mathbf{x}_T \}$ and speaker profiles $\mathbf{D} = \{ \mathbf{d}_1, \dots, \mathbf{d}_K \}$, where $T$ is the length of the acoustic feature, 
and $K$ is the number of the speaker profiles.
The model outputs the multi-speaker transcription 
$\mathbf{Y} = \{y_1, \dots, y_N \}$
and the corresponding speaker labels $\mathbf{S} = \{s_1, \dots, s_N \}$ for each output token, where $N$ is the output length. 
Following the idea of 
serialized output training (SOT) \cite{Serialized-Kanda2020}, 
the multi-speaker transcription $\mathbf{Y}$ is represented by concatenating 
individual speakers' transcriptions
interleaved by a special symbol 
$\langle sc\rangle$
which represents the speaker change.

The E2E SA-ASR consists of two interdependent blocks: 
an {\it ASR block} and a {\it speaker identification block}.
The ASR block 
is represented as follows.
\begin{align}
    \mathbf{E}^{\text{enc}} &= \text{AsrEncoder}(\mathbf{X}), \label{eq:asrenc} \\
    \mathbf{c}_n, \mathbf{\alpha}_n &= \text{Attention}(\mathbf{u}_n, \mathbf{\alpha}_{n-1}, \mathbf{E}^{\text{enc}}), \label{eq:att1} \\
    \mathbf{u}_n &= \text{DecoderRNN}(y_{n-1}, \mathbf{c}_{n-1}, \mathbf{u}_{n-1}), \label{eq:decrnn} \\
    \mathbf{o}_n &= \text{DecoderOut}(\mathbf{c}_n, \mathbf{u}_n, \Bar{\mathbf{d}}_n). \label{eq:decout} 
\end{align}
Firstly, $\text{AsrEncoder}$ maps $\mathbf{X}$ into a sequence of hidden representations, $\mathbf{E}^{\text{enc}}=\left\{ h_1^{\text{enc}}, \dots, h_T^{\text{enc}} \right\}$ (Eq. \eqref{eq:asrenc}). 
Then, an attention module computes  attention weights $\mathbf{\alpha}_n = \left\{ \alpha_{n,1}, \dots, \alpha_{n,T} \right\}$ and context vector $\mathbf{c}_n$ as an attention weighted average of $\mathbf{E}^{\text{enc}}$ (Eq. \eqref{eq:att1}).
The DecoderRNN then computes hidden state vector $\mathbf{u}_n$ (Eq. \eqref{eq:decrnn}). 
Finally, DecoderOut computes 
 the distribution of output token $o_n$ based on the context vector $\mathbf{c}_n$, 
 the decoder state vector $\mathbf{u}_n$,
 and the weighted average of the speaker profiles $\Bar{\mathbf{d}}_n$ (Eq. \eqref{eq:decout}).
Note that $\Bar{\mathbf{d}}_n$ is computed in the speaker identification block, 
which is explained in the next paragraph.
The posterior probability
of token $i$ (i.e. $i$-th token in the dictionary) at the $n$-th decoder step is represented as
 \begin{align}
P(y_n=i|y_{1:n-1},s_{1:n},\mathbf{X},\mathbf{D}) \sim o_{n,i}, \label{eq:tokenprob}
\end{align}
where $o_{n,i}$ represents
the $i$-th element of $\mathbf{o}_n$.

Meanwhile,  
the speaker identification block works as follows.
\begin{align}
    \mathbf{E}^{\text{spk}} &= \text{SpeakerEncoder}(\mathbf{X}), \label{eq:spkenc} \\
    \mathbf{p}_n &= \sum\nolimits_{t=1}^T \mathbf{\alpha}_{n,t} \mathbf{h}_t^{\text{spk}}, \label{eq:spkvec}\\
    \mathbf{q}_n &= \text{SpeakerQueryRNN}(y_{n-1}, \mathbf{p}_n, \mathbf{q}_{n-1}), \label{eq:spkquery}\\
    \mathbf{\beta_{n}} &= \text{InventoryAttention}(\mathbf{q}_n, \mathbf{D}) \label{eq:beta}, \\
    \Bar{\mathbf{d}}_n &= \sum\nolimits_{k=1}^K \beta_{n,k} \mathbf{d}_k. \label{eq:weighted-spk}
\end{align}
Firstly, $\text{SpeakerEncoder}$ converts the input $\mathbf{X}$ to a sequence of hidden representations, $\mathbf{E}^{\text{spk}}=\left\{ h_1^{\text{spk}}, \dots, h_T^{\text{spk}} \right\}$, as speaker embeddings (Eq. \eqref{eq:spkenc}). 
Then, by using the attention weight $\alpha_n$, 
a speaker context vector $\mathbf{p}_n$ is computed for each output token (Eq. \eqref{eq:spkvec}). 
The SpeakerQueryRNN module then generates a speaker query $\mathbf{q}_n$ 
by taking $\mathbf{p}_n$ as an input (Eq. \eqref{eq:spkquery}). 
After getting the speaker query $\mathbf{q}_n$, an InventoryAttention module estimates the attention weights $\mathbf{\mathbf{\beta}}_{n}=\{\beta_{n,1},\dots,\beta_{n,K}\}$ for each speaker profile $\mathbf{d}_k$ in $\mathcal{D}$
(Eq. \eqref{eq:beta}).
Finally,
$\Bar{\mathbf{d}}_n$ is obtained by calculating  
the weighted sum of the speaker profiles using $\beta_n$ (Eq. \eqref{eq:weighted-spk}),
which is input to the ASR block.
In the formulation, the attention weight $\beta_{n,k}$ can be seen as 
a posterior probability of person $k$ speaking the $n$-th token
given all the previous tokens and speakers as well as $X$ and $\mathcal{D}$, i.e., 
\begin{align}
P(s_n=k|y_{1:n-1},s_{1:n-1},X,\mathcal{D}) \sim\beta_{n,k}. \label{eq:spk-prob}
\end{align}

With these model components, all the E2E SA-ASR parameters are trained
by maximizing $\log P(\mathbf{Y},\mathbf{S}|\mathbf{X},\mathbf{D})$, which is defined as 
\begin{align}
\log P(\mathbf{Y},\mathbf{S}|\mathbf{X},\mathbf{D}) \label{eq:samll-1}=&\log\prod_{n=1}^{N}\{P(y_{n}|y_{1:n-1}, s_{1:n}, \mathbf{X}, \mathbf{D}) \nonumber \\ 
&\;\;\;\cdot P(s_{n}|y_{1:n-1}, s_{1:n-1}, \mathbf{X}, \mathbf{D})^\gamma \}, \nonumber
\end{align}
where $\gamma$ is a scaling parameter. 
Decoding with the E2E SA-ASR model is conducted by an extended beam search.
Refer to \cite{Joint-Kanda2020} for further details.

\subsection{Overlapping Inference}
\label{sec:overlap-infer}
Overlapping inference was proposed 
for conventional single-speaker ASR systems
to deal with long-form speech \cite{Comparison-Chiu2019}. 
With the overlapping inference, the input audio is first broken into fixed-length segments. There is $50\%$ overlap between every two consecutive segments so that the information loss around the segment boundaries can be recovered from the overlapping counterpart. 
Given the word hypothesis $\hat{\mathbf{Y}}^m = \{ y_1^m, \dots, y_{N_m}^m \}$, where $y_i^j$ represents the $i$-th recognized word in the $j$-th segment and $N_m$ is the hypothesis length of segment $m$, we can generate two hypothesis sequences by concatenating all the odd segments or even segments as follows:
\begin{align}
    \hat{\mathbf{Y}}^{\text{o}} &= \dots, y_1^{2m-1}, \dots, y_{N_{2m-1}}^{2m-1}, y_1^{2m+1}, \dots, y_{N_{2m+1}}^{2m+1}, \dots, \nonumber \\
    \hat{\mathbf{Y}}^{\text{e}} &= \dots, y_1^{2m}, \dots, y_{N_{2m}}^{2m}, y_1^{2m+2}, \dots, y_{N_{2m+2}}^{2m+2}, \dots. \nonumber
\end{align}
An edit-distance-based algorithm is then applied to align the odd and even sequences, $\hat{\mathbf{Y}}^{\text{o}}$ and $\hat{\mathbf{Y}}^{\text{e}}$. 

To avoid aligning words 
from non-overlapped segments (e.g. segments $m$ and $m+2$),
the distance
between two words from non-overlapped segments is 
set to infinity. 
As a result,
a sequence of word pairs $\langle o_1, e_1 \rangle, \langle o_2, e_2 \rangle, \dots, \langle o_L, e_L \rangle$ is generated:
\begin{align}
    \langle o_i, e_i \rangle = \left\{\begin{array}{ll}
    \langle y^o_{p_o,q_o}, y^e_{p_e,q_e} \rangle & \text{if } y^o_{p_o,q_o} \text{ aligned with }y^e_{p_e,q_e} \\
    \langle \varnothing, y^e_{p_e,q_e} \rangle & \text{if } y^e_{p_e,q_e} \text{ has no alignment} \\
    \langle y^o_{p_o,q_o}, \varnothing \rangle & \text{if } y^o_{p_o,q_o} \text{ has no alignment},
    \end{array} \right. \nonumber
\end{align}
where $i$ is the pair index, $L$ is the total number of matched pairs, $y^o_{p_o,q_o}$ is the $p_{o}$-th word from $q_{o}$-th segment in $\hat{\mathbf{Y}}^{\text{o}}$, $y^e_{p_e,q_e}$ is the $p_{e}$-th word from $q_{e}$-th segment in $\hat{\mathbf{Y}}^{\text{e}}$ and $\varnothing$ denotes no predictions.

The final hypothesis is formed by selecting words from the alignment according to the confidence:
\begin{align}
    \hat{\mathbf{Y}}_i^* = \left\{ \begin{array}{ll}
        o_i & \text{if } f(o_i) \ge f(e_i) \\
        e_i & \text{otherwise.}
    \end{array} \right. \nonumber
\end{align}
Here, $f(\cdot)$ denotes the function to compute the confidence value of
the token.
In \cite{Comparison-Chiu2019},
a simple heuristic of setting a lower confidence value to the
word near the edges of each segment (and a higher confidence value to the word near the center of each segment) was
proposed and shown to be effective.
In this approach,
the confidence score for word $n$ from segment $m$, 
$y_n^m$, is defined as $f(y_n^m) = -|n/C_m - 1/2|$, where $C_m$ denotes the number of words in window $m$.

\subsection{Overlapping inference with E2E SA-ASR}
\label{sec:overlapping-inference-with-sa-asr}

As a baseline system, we consider applying the overlapping inference to the SA-ASR task. 
Because the E2E SA-ASR model generates multiple speakers' transcriptions from each segment, we 
simply apply the overlapping inference for each speaker
independently.
Namely, we first
group the hypotheses from each segment
based on the speaker identities estimated by the E2E SA-ASR model.
We then apply the overlapping inference 
for each speaker's hypotheses.

There are two possible problems
in this procedure.
Firstly, 
speaker misrecognition made by the model could cause a confusion in the alignment process.
Secondly,
even when two hypotheses are observed from 
overlapping segments for the same speaker, 
there is a case where we should not merge the two hypotheses.
This case happens when speakers A and B speak in the order of A-B-A 
and the audio is broken into two overlapped segments such that
the overlapped region contains only speaker B's speech. In this case, we observe  hypotheses of ``A-B'' for the first segment, and hypotheses of ``B-A'' for the second segment. We should not merge the two hypotheses for speaker A because they are not overlapped.
However, avoiding these problems is not trivial.

\section{Hypothesis Stitcher}
\label{sec:stitcher}

\subsection{Overview}
\label{ssec:stitcher-problem}
To handle the long-form multi-talker recordings more effectively, 
we
propose a new method 
using a sequence-to-sequence model, called hypothesis stitcher.
The hypothesis stitcher
consolidates multiple hypotheses obtained from short audio segments into a single coherent hypothesis.
In our proposed approach,
the hypothesis stitcher takes
 the multiple hypotheses of the $k$-th speaker,
$\hat{\mathbf{Y}}^k = \{ \hat{\mathbf{Y}}^{1,k}, \dots, \hat{\mathbf{Y}}^{M,k} \}$, 
where
 $\hat{\mathbf{Y}}^{m,k}$ represents the hypothesis of speaker $k$ for audio segment $m$.
Then, the hypothesis stitcher outputs
a fused single hypotheses 
$\mathbf{H}^k$ given the input $\hat{\mathbf{Y}}^k$.
There are several possible architectures of the hypothesis stitcher,
and we will explain them in the next subsections.

The entire procedure of applying the hypothesis stitcher is as follows.
Firstly, as with the overlapping inference,
a long multi-talker recording is broken up into 
$M$ fixed-length segments with overlaps. 
Then, the SA-ASR model is applied to each segment $m$ to estimate 
the hypotheses $\{ \hat{\mathbf{Y}}^{m,1}, \dots, \hat{\mathbf{Y}}^{m,K} \}$,
where $K$ is the number of the profiles in the speaker inventory $\mathbf{D}$.
After performing the E2E SA-ASR for all segments,
the hypotheses are grouped by the speaker to form 
$\hat{\mathbf{Y}}^k$. 
Here, if speaker $k$ is not detected in segment $m$,
$\hat{\mathbf{Y}}^{m,k}$ is set to empty.
Finally, the hypothesis stitcher works for each speaker
to estimate 
a fused hypothesis $\mathbf{H}^k$.

In this procedure, the input $\hat{\mathbf{Y}}^k$ could have
the same problems 
that we discussed 
for the overlapping inference in Section \ref{sec:overlapping-inference-with-sa-asr}.
However, we expect that the sequence-to-sequence model can work more robustly 
if it is trained appropriately.
In addition,
as an important difference from the overlapping inference, which requires the adjacent segments to overlap by 50\%,\footnote{We noticed a recent study that used the time-alignment information to reduce
the overlap of segments \cite{RNN-Chiu2020}.
However, it requires precise time-alignment information, and is not necessarily be applicable for all systems.}
variants of the hypothesis stitcher can work
on segments with less than 50\% of overlap.
It is important because the smaller the overlap is, the lower the decoding cost becomes.
In the next sections, we will explain the variants of the hypothesis stitcher that we examined in our experiments.

\subsection{Alignment-based stitcher}
\label{ssec:parallel-stitcher}

We propose
an {\it alignment-based stitcher} as an extension of the overlapping inference.
In this method,
the input audio is segmented by a sliding window with 50\% overlap as with the overlapping inference.
Then, the hypotheses from the odd-numbered segments and even-numbered segments
are each joined to yield two sequences as
\begin{align}
\hat{\mathbf{Y}}^{o,k}_{\text{wc}} &= \{ \hat{\mathbf{Y}}^{1,k}, \langle \text{WC} \rangle, \hat{\mathbf{Y}}^{3,k}, \langle \text{WC} \rangle, \dots, \hat{\mathbf{Y}}^{2m-1,k}, \langle \text{WC} \rangle, \dots \}, \nonumber \\
\hat{\mathbf{Y}}^{e,k}_{\text{wc}} &= \{ \hat{\mathbf{Y}}^{2,k}, \langle \text{WC} \rangle, \hat{\mathbf{Y}}^{4,k}, \langle \text{WC} \rangle, \dots, \hat{\mathbf{Y}}^{2m,k}, \langle \text{WC} \rangle, \dots \}. \nonumber
\end{align}
Here, 
we introduce a special window change symbols, $\langle \text{WC} \rangle$, 
to indicate the boundary of each segment in the concatenated hypotheses.
Next, by using the same algorithm as the overlapping inference,
we align $\hat{\mathbf{Y}}^{o,k}_{\text{wc}}$ and $\hat{\mathbf{Y}}^{e,k}_{\text{wc}}$
to generate 
a sequence of word pairs $\langle o_1, e_1 \rangle, \langle o_2, e_2 \rangle, \dots, \langle o_L, e_L \rangle$,
where $o_l$ and $e_l$ can be $\langle \text{WC} \rangle$.
Finally, this word pair sequence is
input to the hypothesis stitcher model to estimate the fused single hypothesis $\mathbf{H}^k$.
In this paper, we 
represent the hypothesis stitcher by 
 the transformer-based attention encoder-decoder \cite{vaswani2017attention},
and we simply concatenate the embeddings of $o_l$ and $e_l$ for each position $l$ to 
form the input to the encoder.

In the overlapping inference,
the word-position-based confidence function $f(\cdot)$ is used to select the word from the aligned word pairs.
On the other hand,
in the alignment-based stitcher,
we expect 
better word selection to be performed by the sequence-to-sequence model 
by training the model on 
an appropriate data.

\subsection{Serialized stitcher}
\label{ssec:serialized-stitcher}

We also propose another architecture, called a {\it serialized stitcher}, which 
is found to be effective while being much simpler than the alingment-based stitcher.
In the serialized stitcher, 
we simply join all hypotheses of the $k$-th speaker from every short segments as
\begin{align}
    \hat{\mathbf{Y}}^k_{\text{wc}} = \{ \hat{\mathbf{Y}}^{1,k}, \langle \text{WC} \rangle, \hat{\mathbf{Y}}^{2,k}, \langle \text{WC} \rangle, \hat{\mathbf{Y}}^{3,k}, \langle \text{WC} \rangle, \dots, \hat{\mathbf{Y}}^{M,k} \}. \nonumber
\end{align}
Then, $\hat{\mathbf{Y}}^k_{\text{wc}}$ is fed into the hypothesis stitcher to estimate
the fused single hypothesis.
We again used the transformer-based attention encoder-decoder to represent the hypothesis stitcher.

We also examine a variant of the serialized stitcher where we insert 
 two different symbols $\langle \text{WCO} \rangle \text{ and } \langle \text{WCE} \rangle$ to explicitly indicate
 the odd-numbered segment and even-numbered segment.
\begin{align}
    \hat{\mathbf{Y}}^k_{\text{wco/e}} = \{ \hat{\mathbf{Y}}^{1,k}, \langle \text{WCO} \rangle, \hat{\mathbf{Y}}^{2,k}, \langle \text{WCE} \rangle, \hat{\mathbf{Y}}^{3,k}, \langle \text{WCO} \rangle, \dots, \hat{\mathbf{Y}}^{M,k} \}. \nonumber
\end{align}

Note that the serialized stitcher no longer requires
the short segments to be 50\% overlapped because there is no alignment procedure.
In Section \ref{sec:exp}, 
we examine how the overlap ratio between the adjacent segments affects the accuracy of the serialized stitcher.

\section{Experiments}
\label{sec:exp}
We conduct a basic evaluation
of the hypothesis stitcher 
by using simulated mixtures of the LibriSpeech utterances  \cite{Librispeech-Panayotov2015}
in Section \ref{ssec:eval-on-librispeech}.
The best model is then evaluated 
on the LibriCSS \cite{Continuous-Chen2020}, a noisy multi-talker audio
recorded in a real meeting room, 
in Section \ref{ssec:res-libricss}.

\subsection{Evaluation with clean simulated mixture of LibriSpeech}
\label{ssec:eval-on-librispeech}

\subsubsection{Evaluation settings}
\label{ssec:eval-set}

Our experiments used the E2E SA-ASR model of \cite{Investigation-Kanda2020}, which was 
trained using LibriSpeech.  
The training data  were generated by 
randomly mixing multiple utterances in ``train\_960''.
Up to 5 utterances were mixed for each sample, and the average duration was 29.4 sec.
Refer to \cite{Investigation-Kanda2020} for further details of the model and training configurations. 

On top of the E2E SA-ASR model, 
we trained hypothesis stitcher models by using a  16-second-long sliding window.
We first simulated monaural multi-talker data 
from LibriSpeech (see  Table~\ref{tab:data}).
There were $50,083$ long-form audio training samples, 
each of which was a mixture of multiple utterances randomly selected from ``train\_960''.
Each sample consisted of up to 12 utterances spoken by 6 or fewer speakers. 
When the utterances were mixed, 
each utterance was shifted by a random delay to simulate partially overlapped conversational speech so that an average overlap time ratio became $10\%$.
All training samples were then broken into 16-second overlapped segments in a simlar way to the settings of the overlapping inference paper \cite{Comparison-Chiu2019}, 
and they were decoded by the E2E SA-ASR model 
with relevant speaker profiles.
The generated hypotheses and the original reference label were used for the input and output, respectively, to train
the hypothesis stitcher. 

For all variants of the hypothesis stitchers, we used
the transformer-based attention encoder-decoder with 6-layers of encoder (1,024-dim, 8-heads) and 6-layers of decoder (1,024-dim, 8-heads)
by following \cite{vaswani2017attention}. The model was trained by using an Adam optimizer with a learning rate of 0.0005
until no improvement was observed for 3 consecutive epochs on the development set.
We applied the label smoothing and dropout with parameters of 0.1 and 0.3, respectively.

For evaluation, we generated two types of development and test sets:
a short set with roughly 30-second-long samples and a long set with roughly 60-second-long samples (Table \ref{tab:data}).
We used the SA-WER, 
which was calculated by comparing the ASR hypothesis and the reference transcription of each speaker, as an evaluation metric.
\begin{table}[t]
    \centering
    \scriptsize
    \caption{Summary of the clean simulated data set}
    \label{tab:data}
    \begin{tabular}{l|c|c|c|c|c}
        \toprule
           & Train & \multicolumn{2}{c|}{Dev} & \multicolumn{2}{c}{Test} \\
           &       & short & long & short & long \\
        \midrule
     Source  & train\_960 & dev\_clean & dev\_clean & test\_clean & test\_clean  \\
    \# of samples      & 50,083 & 513        & 319        & 496   & 310                      \\
    Avg. words. & 193 & 106     & 171        & 106   & 170                 \\
    Avg. dur. (sec) & 61.9        & 34.1         & 54.1    & 35.4    & 56.1                  \\ \midrule
    Total dur. (hr)      & 861.6 & 4.9        & 4.8        & 4.9   & 4.8                      \\
    \bottomrule
    \end{tabular}
\vspace{-5mm}
\end{table}

\begin{table}[t]
\centering
\caption{SA-WER (\%) on clean simulated evaluation set.}
\footnotesize
\begin{tabular}{l|c|cc|cc}
\toprule
Decoding method & Segment &\multicolumn{2}{c|}{Dev} & \multicolumn{2}{c}{Test} \\
& overlap  &  short & long & short & long \\
\midrule
 No segmentation &- & 14.2 & 17.2 & 13.1 & 21.1   \\
 Block-wise inference &0\%    & 11.8 & 12.4 & 12.5 & 15.2   \\
 Overlapping inference&50\%  & 12.0 & 12.8 & 12.7 & 15.2   \\
\midrule
 Stitcher (alignment-based)&50\% & 11.4 & 12.0 & 12.3 & 14.5 \\
Stitcher (serialized, WC) &50\%   & 10.6 & 11.9 & 10.8 & 13.7   \\
Stitcher (serialized, WCO/E) &50\%   & \textbf{10.5} & \textbf{11.5} & \textbf{10.6} & \textbf{13.4}   \\
\bottomrule
\end{tabular}
\label{tab:res-simlibricss}
\vspace{-5mm}
\end{table}

\subsubsection{Results}

First, we evaluated the hypothesis stitcher using a sliding window of 50\% overlap. 
Table~\ref{tab:res-simlibricss} shows the results. 
As the baseline, we evaluated the E2E SA-WER without segmentation,
denoted as
``No segmentation'' in the table.
We also evaluated a ``Block-wise inference'' method, in which the E2E SA-ASR was performed by using a non-overlapping 16-second sliding window. For this method, the estimated hypotheses from each window were naively concatenated.
As shown in the table, we observed a significant improvement by this simple method
especially on the long evaluation set. 
This result suggests that the E2E SA-ASR that was used for
the evaluation did not generalize well for very long audio.
We then evaluated the overlapping inference using 16 seconds of segments with 8 seconds of segment shift.
However, the overlapping inference showed
slightly worse results
than the simple block-wise inference. 
Note that, in our preliminary experiment,
we have confirmed that the overlapping inference
achieved better accuracy than the block-wise inference
when the recording consists of only one speaker.
Therefore, the degradation could have been caused by the two types of issues as we discussed in Section \ref{sec:overlapping-inference-with-sa-asr}.
Finally, the proposed hypothesis stitcher models were evaluated.
We found that all variants of the hypothesis stithcers significantly outperformed all the baseline methods.
We also found that the serialized stitcher with the two types of auxiliary symbols, $\langle \text{WCO} \rangle \text{ and } \langle \text{WCE} \rangle$, worked the best.

Given the serialized stitcher yielding the best performance, we further conducted the evaluation 
by using a sliding window with a less overlap ratio (i.e. with a larger stride of the window). 
The results are shown in Table~\ref{tab:res-overlaps}.
While the best performance was still obtained with the half-overlapping segments, 
the serialized stitcher worked well 
even with the non-overlapping segments.
In practice, the segment overlap ratio could be chosen by considering the balance
between the computational cost and accuracy.

\begin{table}[t]
\centering
\caption{SA-WER (\%) with various overlap ratio of segments.}
\footnotesize
\begin{tabular}{l|c|cc|cc}
\toprule
Decoding method   & Segment & \multicolumn{2}{c|}{Dev} &\multicolumn{2}{c}{Test} \\
          & overlap & short & long & short & long \\
\midrule
Stitcher (serialized, WCO/E) & 0\% & 11.6 & 12.4 & 12.3 & 15.1 \\
Stitcher (serialized, WCO/E) & 25\% & 11.0 & 11.9 & 11.5 & 14.3 \\ 
Stitcher (serialized, WCO/E) & 50\% & 10.5 & 11.5 & 10.6 & 13.4 \\
\bottomrule
\end{tabular}
\label{tab:res-overlaps}
\vspace{-6mm}
\end{table}

\subsection{Evaluation with LibriCSS corpus}
\label{ssec:res-libricss}

\begin{table}[t]
\centering
\caption{SA-WER (\%) on LibriCSS.}
\scriptsize
\begin{tabular}{l|cccccc|c}
\toprule
Method                 & \multicolumn{6}{c|}{Speech overlap ratio (\%)} & Total\\
                 & 0S  & 0L  & 10 & 20 & 30 & 40 &   \\
\midrule
No segmentation & \textbf{6.9} & \textbf{7.0} & 11.2 & 15.0   & 28.4 & 30.3 & 17.8 \\
Stitcher (serialized, WCO/E)    & 7.0 & 7.6 & \textbf{10.4} & \textbf{14.6} & \textbf{23.2} & \textbf{25.4} & \textbf{15.7}\\
\bottomrule
\end{tabular}
\label{tab:res-libricss}
\vspace{-5mm}
\end{table}

Finally, we 
evaluated the proposed method
on the LibriCSS dataset \cite{Continuous-Chen2020}.
The dataset consists of 10 hours of recordings of concatenated 
LibriSpeech utterances that were played back by 
multiple loudspeakers in a meeting room 
and captured by a seven-channel microphone array. 
We used only the first channel data (i.e. monaural audio) for our experiments.
Before applying the decoding procedure,
we split the recordings at every silence regions by an oracle voice activity detector (VAD)
that uses the reference time information.
Note that the audio set after VAD still contains
considerable amount of long-form multi-talker audio.
Each recording consists of utterances of 8 speakers. 
When we applied the E2E SA-ASR,
we fed the speaker profiles 
corresponding to the 8 speakers for each recording.

We used the serialized stitcher
 trained in the previous experiments on the segments obtained by a 50\% overlapping sliding window.
The evaluation result is shown in Table~\ref{tab:res-libricss}.
According to the convention in \cite{Continuous-Chen2020},
we report the results for 
different overlap speech ratios
ranging from 0\% to 40\%.
As shown in the table, we observed a large improvement when 
the overlap speech ratio was large (30--40\%).
This is because these test sets contained much more long-form audio 
due to the frequent utterance overlaps.





\section{Conclusion}
\label{sec:conclusion}

In this paper, we proposed the hypothesis stitcher 
to help improve the accuracy of the E2E SA-ASR model on the long-form audio. 
We proposed several variants of the model architectures for the hypothesis stitcher. The experimental results showed that one of the model called serialized stitcher worked the best.
We also showed that the hypothesis stitcher yielded good performance even with sliding windows with less than 50\% overlaps, 
which is desirable to reduce the computational cost of the decoding.

\bibliographystyle{IEEEtran}
\bibliography{strings,refs,additional}

\end{document}